\pdfoutput=1
\documentclass[aps,prl,twocolumn,showpacs,superscriptaddress,a4paper,floatfix]{revtex4}

\usepackage[latin1]{inputenc}
\usepackage{siunitx}
\usepackage{hyperref}
\usepackage{datetime}
\usepackage{graphicx}
\usepackage{textcomp}
\usepackage{microtype}
\usepackage{amsmath,amssymb}

\begin{document}

\title{A scanning gate microscope for cold atomic gases}

\author{Samuel Häusler}
\affiliation{Department of Physics, ETH Zurich, 8093 Zurich, Switzerland}
\author{Shuta Nakajima}
\affiliation{Department of Physics, Graduate School of Science, Kyoto University, Kyoto 606-8502, Japan}
\author{Martin Lebrat}
\author{Dominik Husmann}
\author{Sebastian Krinner}
\author{Tilman Esslinger}
\affiliation{Department of Physics, ETH Zurich, 8093 Zurich, Switzerland}
\author{Jean-Philippe Brantut}
\affiliation{Institute of Physics, EPFL, 1015 Lausanne, Switzerland}

\date{\pdfdate}

\begin{abstract}
We present a scanning probe microscopy technique for spatially resolving transport in cold atomic gases, in close analogy with scanning gate microscopy in semiconductor physics. The conductance of a quantum point contact connected to two atomic reservoirs is measured in the presence of a tightly focused laser beam acting as a local perturbation that can be precisely positioned in space. By scanning its position and recording the subsequent variations of conductance, we retrieve a high-resolution map of transport through a quantum point contact. We demonstrate a spatial resolution comparable to the extent of the transverse wave function of the atoms inside the channel, and a position sensitivity below \SI{10}{\nano\metre}. Our measurements agree well with an analytical model and \emph{ab-initio} numerical simulations, allowing us to identify a regime in transport where tunneling dominates over thermal effects. Our technique opens new perspectives for the high-resolution observation and manipulation of cold atomic gases.
\end{abstract}

\pacs{
}

\maketitle
Scanning probe microscopes had substantial impact on the development of solid-state physics during the last three decades, from the observation of individual atoms at surfaces \cite{RevModPhys.59.615,RevModPhys.75.949}, to the imaging of coherent electron flow \cite{Topinka:2001aa} and the identification of order parameters in complex correlated materials \cite{Fischer:2007aa, Allan2015} -- just to name a few examples. Many of these applications rely on two conceptually important ingredients: (i) the use of very sharp probes positioned with atomic-scale precision close to a surface, (ii) the ability to continuously measure transport in the presence of a local probe, which yields precise information related to a single point of the system by accumulating the often weak transport signal.

Many fundamental phenomena observed in condensed matter physics are also studied in cold-atoms based quantum simulations. This has motivated the development of high spatial resolution imaging based on photon \cite{Gemelke:2009aa,Bakr:2009aa,Sherson:2010aa,PhysRevLett.114.213002,PhysRevLett.114.193001,Haller:2015aa,PhysRevA.92.063406,PhysRevLett.115.263001,Yamamoto:2015aa,PhysRevLett.116.175301} and electron \cite{Gericke:2008aa} scattering, which typically yields a destructive observation of the local density distribution or the parity of the atom number on lattice sites. Yet, a high spatial resolution measurement in a transport setting, as has been so successfully applied in solid-state physics, has so far not been carried out with atomic gases. 

In this letter, we demonstrate a scanning gate microscope for a cold atomic gas flowing through an optically created quantum point contact (QPC) \cite{Krinner:2015aa}. Our technique is inspired by scanning gate microscopy in semiconductor physics \cite{Eriksson:1996aa,Topinka:2000aa,Topinka:2001aa,Sellier:2011aa}, where a movable gate potential is used to locally modify the underlying carrier density in a sample. In our cold-atom implementation, we use a high-resolution optical microscope to create a sub-micrometer repulsive gate potential in the region of the QPC. Thanks to the intrinsic diluteness of cold atomic gases, our gate operates at the scale of the Fermi wavelength.

\begin{figure}
    \includegraphics[width=0.5\textwidth]{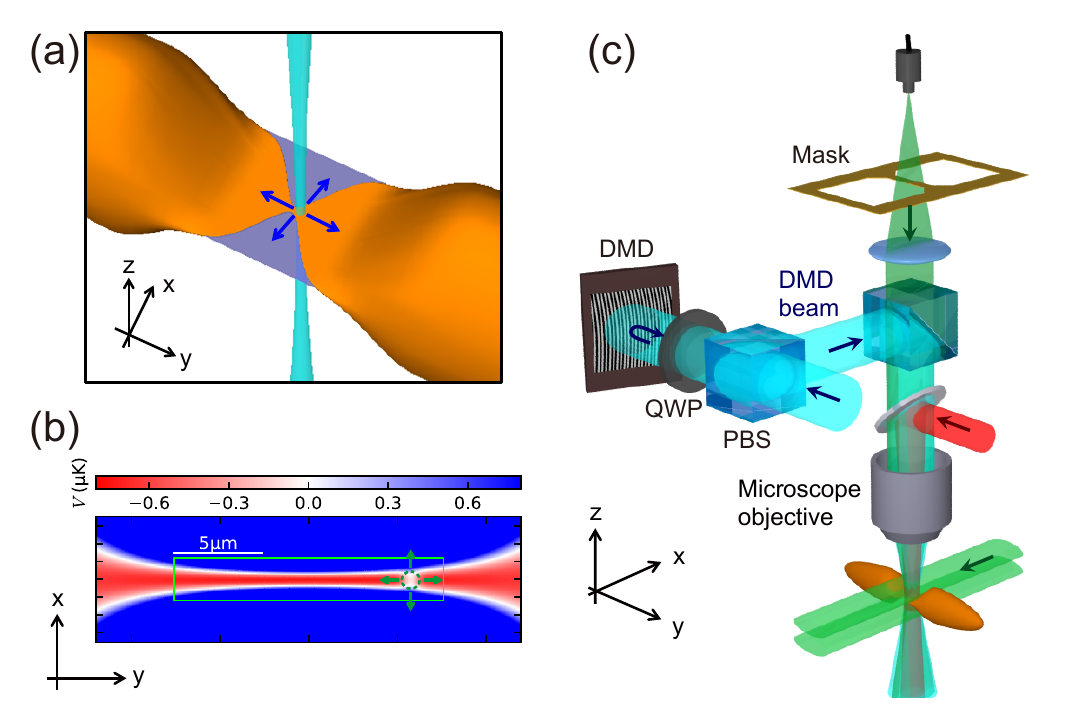}
    \caption{An atomic scanning gate microscope. (a) Close-up view of the channel region: mesoscopic reservoirs, 2D regions and the QPC. The scanning gate is realized by a tightly focused, repulsive laser beam that is scanned across the structure. (b) The gate (green circle) locally modifies the potential landscape shown for the parameters used in the simulation, panel \ref{fig:2dscan}b. The green square indicates the region mapped in figure~\ref{fig:2dscan}. (c) Engineering of the QPC and the scanning gate. The QPC is formed at the intersection of two repulsive laser beams (indicated in green) having a nodal line at the center. A TEM$_{01}$-like laser mode propagating along the $-x$ axis forms the 2D regions. The beam passing through a lithographic mask is shrunk onto the 2D region and creates the QPC. The scanning gate is holographically shaped and moved (blue beam) by a DMD and projected onto the atoms with a microscope objective. The red beam creates an attractive potential to locally shift the chemical potential.}
    \label{fig:setup}
\end{figure}

Our technique complements the direct fluorescence or absorption imaging in many respects. At the conceptual level, it uses quantum degenerate atoms themselves, rather than photons, as test particles incident on the system \cite{Imry:1999aa}. Large reservoirs connected to a smaller, mesoscopic system act as source and sink for the scattered atoms, continuously accumulating the signal. Since no spontaneous emission of photons or other dissipative processes are induced during the accumulation, it is possible to access long time scales. In contrast, photon or electron scattering provide an instantaneous snapshot of the density distribution.

The basis of our experiment is a quantum degenerate Fermi gas of $^6$Li atoms, as described in our previous work \cite{Krinner:2015aa}. The Fermi gas is produced in a combined magnetic and optical trap, yielding an elongated cloud with $N = \num{1.01(7)e5}$ atoms in each of the lowest and third lowest hyperfine states of lithium. A homogeneous magnetic field of \SI{949}{G} is applied, which sets the scattering length to $-2653\,a_0$, where $a_0$ is Bohr's radius. This corresponds to an interaction parameter $1/k_\text{F}a$ in the reservoirs of \num{-2.1}, where $k_\text{F} = \sqrt{2 m E_\text{F}}/\hbar$ is the Fermi wavevector in the gas, $m$ is the mass of lithium atoms and $E_\text{F} = \hbar \bar{\omega} (6N)^{1/3}$ is the Fermi energy in the harmonic trap, with $\bar{\omega}$ the geometric mean of its frequencies. At typical temperatures of about \SI{60}{\nano\kelvin} we expect our gas to be in the normal phase, as the critical temperature for superfluidity is \SI{42}{\nano\kelvin} \cite{supplement}. As presented in figure \ref{fig:setup}a, a repulsive potential generated by a laser beam with a nodal line in the middle is imposed on the cloud, creating a quasi two-dimensional Fermi gas at the center of the cloud, smoothly connected on both sides to large, three-dimensional reservoirs \cite{Brantut:2012aa}. The trap frequency (mode spacing) along the vertical ($z$) direction at the center of the quasi two-dimensional region reaches $\omega_z = 2\,\pi \cdot \SI{13(5)}{\kilo\hertz}$. The QPC is produced by imaging a binary mask using light at $532\,$nm, imprinting a thin wire onto the quasi two-dimensional region similar to \cite{Krinner:2015aa}. We reach trap frequencies along the transverse direction of about $\omega_x = 2\,\pi \cdot \SI{22(9)}{\kilo\hertz}$ at the center of the QPC. Along the transport direction ($y$), the beam producing the QPC has a waist of \SI{9.1(3)}{\micro\metre}. An attractive potential produced by a Gaussian, red-detuned beam with a waist of \SI{42.5(3)}{\micro\metre} is superimposed onto the QPC, allowing for the control of the chemical potential in the QPC and its immediate vicinity. Upon increasing the chemical potential, successive transverse modes of the QPC are populated yielding characteristic conductance plateaus \cite{Krinner:2015aa}.

\begin{figure}
    \includegraphics{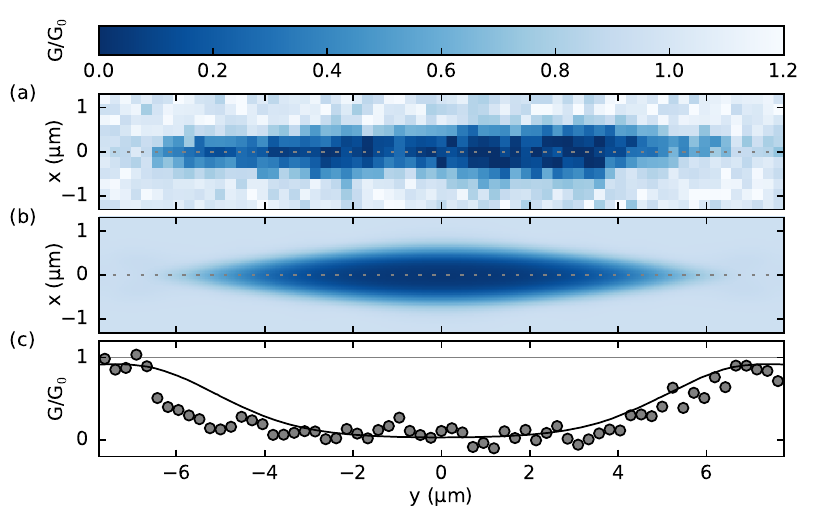}
    \caption{Scanning gate map of a single mode QPC. Measured (a) and calculated (b) conductance~$G$ normalized to the background value $G_0$ as a function of the position of the scanning gate imposed onto the QPC. Due to the weak attraction among the particles, the measured background of $1.3/h$ is larger than the universal conductance of $1/h$ \cite{Krinner:2016}. The scanning gate has a strength of $\SI{620(1)}{\nano\kelvin} \cdot k_\text{B}$. The reservoir acting as source is located on the left and the sink on the right side of the map. The simulation considers a single free parameter, the chemical potential in the reservoirs. The temperature and the chemical potential bias are \SI{58(7)}{\nano\kelvin} and $\SI{99(4)}{\nano\kelvin} \cdot k_\text{B}$ respectively, calibrated independently \cite{supplement}. (c) Longitudinal cuts through the scanning gate maps, along the dashed lines in panel (a) and (b). The measured conductances are represented by the points and the simulation by the line.}
    \label{fig:2dscan}
\end{figure}

The scanning gate potential is produced using light at \SI{532}{\nano\metre} tightly focused onto a spot with waists of $w_x = \SI{731(1)}{\nano\metre}$ and $w_y = \SI{751(1)}{\nano\metre}$. This beam is shaped and controlled using a digital mirror device (DMD), operating in the Fourier plane of the microscope as a diffraction grating in Littrow configuration (see figure \ref{fig:setup} and \cite{supplement}). The discreteness of the DMD sets the minimal displacement to \SI{93}{\nano\metre} with our optical setup.

We operate the QPC in the single mode regime by tuning the chemical potential to the center of the first plateau. This condition, together with the small size of the tip and the symmetry of our potential represents a unique case, where the response of the system to the scanning gate can be interpreted as a map of the current distribution in the weak probe limit \cite{PhysRevB.88.035406}. We scan the position of the gate in the region indicated in figure~\ref{fig:setup}b with an extent of \SI{15.2}{\micro\metre} $\times$ \SI{2.4}{\micro\metre}. The resulting map is shown in figure \ref{fig:2dscan}, where each pixel represents the conductance of the QPC with the scanned gate at the position of the pixel. The individual measurements are separated by \SI{238}{\nano\metre} and taken with a strength of the scanning gate of $V_0 = \SI{620(1)}{\nano\kelvin} \cdot k_\text{B}$. The strength $V_0$ is about twice the local Fermi energy at the center of the structure, corresponding to the strong probe regime \cite{Szewc2013}. The region of low conductance represents the center of the QPC, where the current density is the highest. The pattern fades out towards the edges along $y$, where the current density is smaller due to the weaker confinements. Classically, these are regions where the extension of the gate is smaller than the width of the conductor. The full width at half maximum (FWHM) of the conductance pattern along the $y$ direction is \SI{12(1)}{\micro\metre}, matching that of the beam creating the QPC. Along the transverse ($x$) direction, the short FWHM of \SI{0.8(2)}{\micro\metre} results from the tight confinement of the QPC.

We compare the experimental results with direct numerical simulations of the scanning gate setup using the Kwant library \cite{supplement,Groth:2014}. This solves the scattering problem of independent particles originating from ideal reservoirs and impinging onto the structure. The potential landscape of the QPC along $x$ and $y$ is set \emph{a priori} from the geometry of the laser beams, and the chemical potential is adjusted to fit the data. The results of the simulation are shown in figure \ref{fig:2dscan}b, showing overall good agreement with the experiment. In particular, the transverse and longitudinal shapes are reproduced, as well as the fading out of the pattern in the wings of the QPC. 

It was observed in the condensed matter context that scanning gate maps are dressed by fringe patterns, resulting from interferences between particles emitted by the point contact and reflected by the scanning gate \cite{Topinka:2001aa}. In our experiment, these fringes are washed out by finite temperature, as confirmed by our numerical simulations \cite{supplement}. In contrast to semiconductor nanostructures, where large scale disorder channels the particles emitted by the QPC \cite{Topinka:2001aa}, our system is free of disorder, and channeling does not take place. 

\begin{figure}[htbp]
    \includegraphics{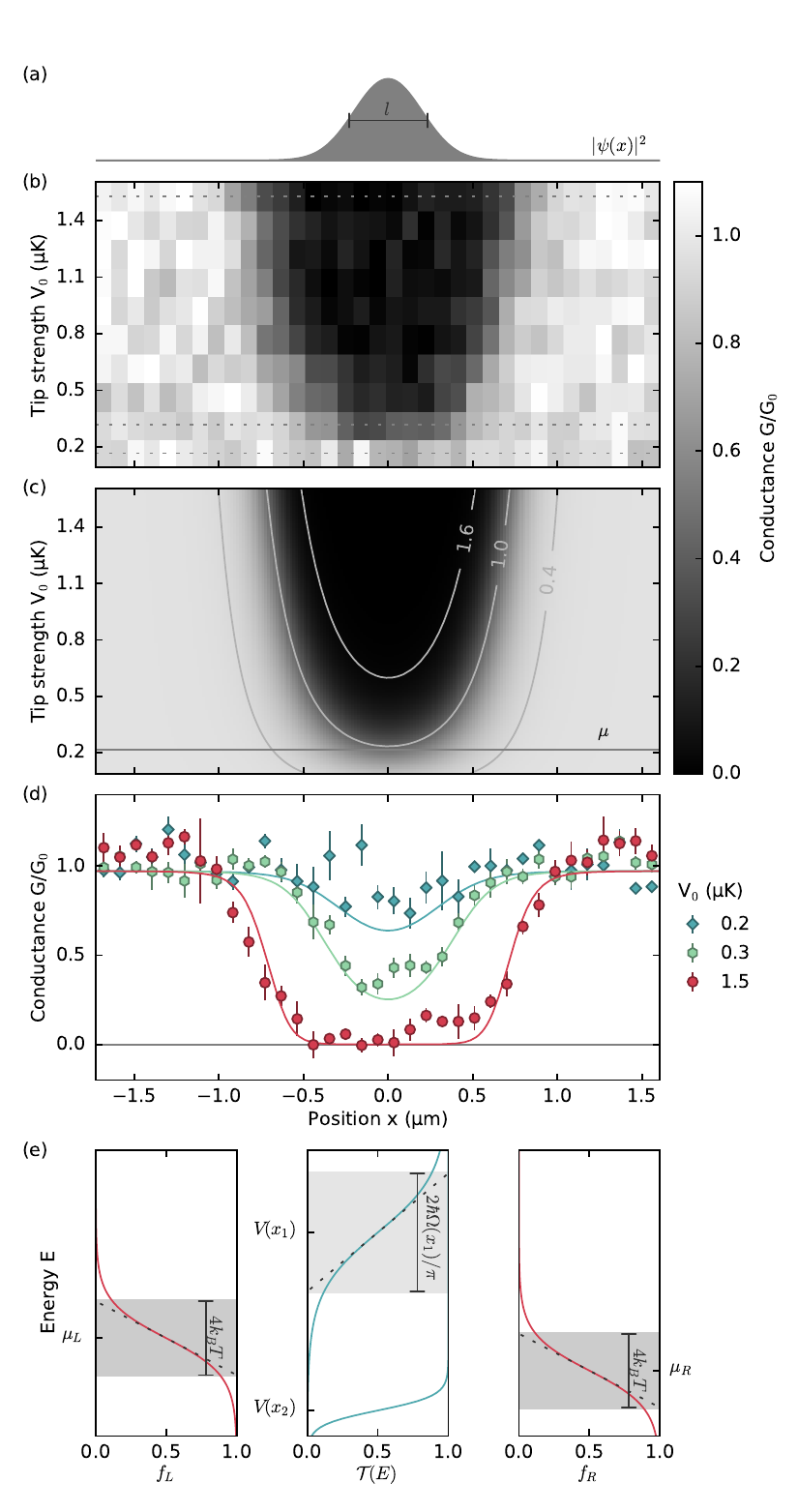}
    \caption{Transverse scans through the QPC's center in the single mode regime. (a) Transverse ground state probability distribution with FWHM~$l$ of \SI{0.5(2)}{\micro\metre} in absence of the tip. (b) Conductance normalized to the measured background value $G_0$ of $1.1/h$ as a function of the tip position~$x$ and strength~$V_0$. The dotted lines indicate representative cuts shown in (d). (c) Prediction by the analytical model including chemical potential as the only free parameter, while temperature and chemical potential bias are independently calibrated to \SI{58(5)}{\nano\kelvin} and $\SI{102(3)}{\nano\kelvin} \cdot k_\text{B}$. The fitted chemical potential~$\mu$ is indicated in the plot. Solid lines mark realisations with the same ratio $\hbar \Omega(x) / 2 \pi k_\text{B} T$. (d) Transverse scans for different tip strengths~$V_0$. The solid curves indicate best fits with the analytical model. (e) Fermi-Dirac distribution~$f_\text{L/R}(E)$ of each reservoir centered around the chemical potential~$\mu_\text{L/R}$. They are smeared over the energy scale~$4 k_\text{B} T$ (dark shaded area), defined by the tangential line (dashed line) at the inflection point. The transmission~$\mathcal{T}(E)$ through a parabolic barrier for two different heights $V(x)$ and the associated energy scale $2\hbar \Omega(x) / \pi$ (light shaded area). The upper (lower) curve is shown in the tunneling (thermal) dominated regime, for a ratio~$\hbar \Omega(x) / 2 \pi k_\text{B} T$ of 1.6 (0.4).}
    \label{fig:cuts}
\end{figure}

We now study the regimes of scanning gate microscopy, from weak to strong probes. To this end, we scan the gate transversally through the center of the QPC, with varying $V_0$. These cuts are shown in figure \ref{fig:cuts}. For the lowest $V_0$, the channel is not closed even with the scanning gate at the center of the QPC. This corresponds to the weak probe regime. As $V_0$ is increased, the conductance quickly goes to zero when the tip is at the center, and the profile changes from approximately Gaussian to flat-top. For stronger scanning gates, the QPC is fully closed over an increasingly wide range, reflecting a clipping effect.

To analytically model the process we assume that the particles only explore a longitudinal cut through the Gaussian tip potential as the transverse wave function is narrower than the tip. The ground state wave function has a FWHM of $l = 1.67 \sqrt{\hbar / m \omega_x} = \SI{0.5(2)}{\micro\metre}$ inside the QPC, about half that of the tip in transverse direction. Furthermore, we approximate the potential cut in transport direction ($y$) by a parabolic barrier with anti-trapping frequency $\Omega (x) = 2/w_y \sqrt{V(x)/m}$ and potential offset $V(x) = V_0 \mathrm{e}^{-2x^2/w_x^2}$, where $x$ is the transverse tip position with respect to the QPC center. The transmission $\mathcal{T}$ through the parabolic barrier is given by 
\begin{equation}
	\mathcal{T}(E) = \frac{1}{1+\mathrm{e}^{2 \pi \frac{E - V(x)}{\hbar \Omega (x)}}}, 
	\label{eq:saddlePoint}
\end{equation}
where $E$ is the energy of the incident particle \cite{Glazman:1988aa,Ihn:2010aa}. We combine the transmission~$\mathcal{T}(E)$ with the thermal occupation of states in the reservoirs to obtain conductances, using Landauer's formula \cite{Ihn:2010aa}. The profiles calculated using this model are shown in figure \ref{fig:cuts}c, where the overall chemical potential is the only free parameter common to all curves. The agreement with the measurement is good over the whole range of parameters, as can be seen on the cuts in figure \ref{fig:cuts}d. We also compared this model with the numerically exact Kwant simulation, finding good agreement \cite{supplement}.

Interestingly, the analytical model allows to distinguish transport based on quantum tunneling through the gate from thermally activated particles in the reservoirs. The Landauer formula includes both contributions and expresses conductance as the convolution of $\mathcal{T}(E)$ with the difference of the Fermi distributions. The model yields a transmission that has the same form as the Fermi distribution, allowing us to quantitatively evaluate the relative roles of tunneling and finite temperature, as illustrated in figure~\ref{fig:cuts}e. We can access regimes where tunneling dominates over thermal effects meaning $\hbar \Omega(x) /2 \pi > k_\text{B} T$ \cite{Affleck1981}, as presented in figure~\ref{fig:cuts}c. This is in strong contrast with experiments in condensed matter physics where direct tunneling through the scanned gate is negligible.

\begin{figure}
    \includegraphics{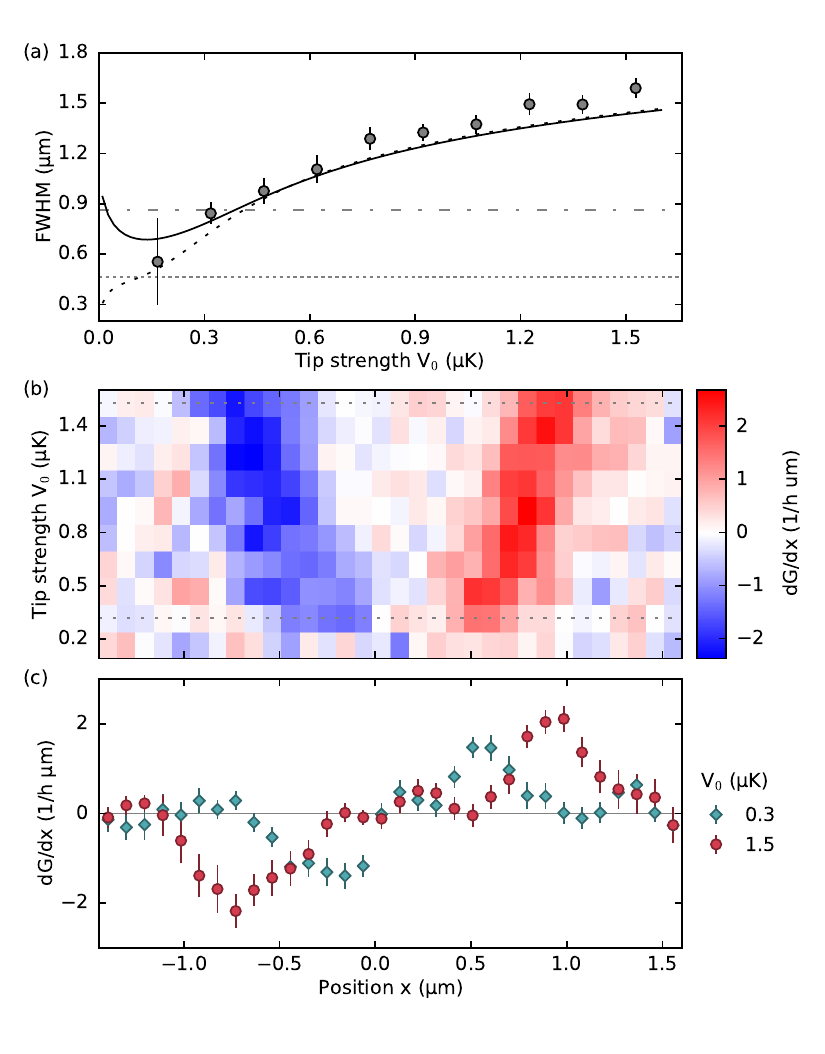}
    \caption{Resolution and sensitivity of the scanning gate technique. (a) FWHM of the transverse scans through the QPC's center, shown in figure~\ref{fig:cuts}, as a function of the tip strength. Beside the experimental points the dashed and solid lines represent the predictions of the analytical model for zero and finite temperature of \SI{58(5)}{\nano\kelvin}. For comparison the FWHM of the obstacle in transverse direction (dashed dotted line) and of the transverse probability distribution (horizontal dotted line), depicted in figure~\ref{fig:cuts}a, are indicated. (b) Derivative of the measured conductance with respect to the position of the tip as a function of its position~$x$ and its strength~$V_0$. (c) Representative cuts along the dashed lines in panel (b). Conductance measurements are maximally sensitive to position, for the strongest gate, where the derivative gets extremal.}
    \label{fig:width}
\end{figure}

The non-linearity of the transmission coefficient in equation~\eqref{eq:saddlePoint} has important consequences for the spatial resolution. To estimate the resolution we measure the FWHM of the transverse cuts of figure \ref{fig:cuts} \cite{supplement}. The results are presented in figure \ref{fig:width}a, as a function of $V_0$. For strong gates the FWHM is large because the QPC is already blocked by the raising edges of the Gaussian gate. Reducing $V_0$, the FWHM goes down and becomes smaller than that of the laser beam for $V_0 < \SI{0.4}{\micro\kelvin}$. Interestingly, it keeps decreasing for lower $V_0$, showing that the resolution is not limited by the optical beam profile, analogous to super-resolved optical techniques reaching resolutions beyond the diffraction limit \cite{Novotny:2012aa}. For the smallest $V_0$, the signal is weak but the FWHM is low enough to be comparable with that of the transverse ground state wave function in the QPC. It agrees with the analytical model, which also predicts very strong thermal broadening for weak scanning gates.

While weak gates allow for high spatial resolution, strong gates maximize the position sensitivity, because small position changes can yield large variations in conductance. This is the case in the raising edges of a strong scanning gate, and is widely exploited in scanning gate microscopy in the solid state context \cite{Sellier:2011aa}. To study this effect, we extract the derivative of conductance with position $\mathrm{d}G / \mathrm{d}x$, as shown in figure \ref{fig:width}b \cite{supplement}. The extremal variation rates mark the falling edges of the profiles and separate with increasing gate strength. The evolution of the width of the profile is clearly visible, as well as the clipping regime when the strong scanning gate is located at the center of the QPC. The fastest variations amount to $2.7(3) / h$ per micrometer.

The position sensitivity of our apparatus is limited by the signal to noise ratio with which we can measure conductances. The figure of merit is $\delta x = (\mathrm{d}G / \mathrm{d} x)^{-1} \delta G$, where $\delta G$ is the noise in the conductance measurement. To assess the minimal noise we use the overlapping Allan deviation \cite{Riley:2008aa}, giving $0.024(12)/h$ \footnote{We performed 128 repeated conductance measurements in the single mode regime without scanning gate. The Allan deviation becomes minimal upon binning 29 consecutive measurements. \cite{supplement}}. This translates into a position sensitivity of $\delta x = \SI{9(5)}{\nano\metre}$. The sensitivity characterizes our instrument, and is mainly limited by the shot-to-shot noise in the preparation of the reservoirs.

Since the scanned gate is optically generated with programmable holograms, the same setup could be used to project several gates, with tailored shapes, serving as building blocks for more complex atomtronics circuits \cite{Albiez:2005aa,Pepino:2009aa,PhysRevA.82.013640,Eckel:2014aa,Husmann:2015ab,Valtolina:2015ab,PhysRevA.93.063619}. Time-modulated or near-resonant optical gates that address external or internal atomic degrees of freedom could allow to locally generate effective gauge structures. Such gates could also be used to perform spectroscopic measurements in analogy with scanning tunneling microscopy \cite{PhysRevLett.115.165301}.

Our scanning gate technique can be generalized to any cold atoms system in which conductance measurements can be performed, such as disordered systems \cite{Shapiro:2012aa}, yielding the additional ability to control the potential at a scale shorter than the localization length. It could distinguish percolation processes from localization by interferences, or be combined with density measurements to identify the fraction of the atoms participating in transport \cite{Krinner:2015ac}. In superfluid Fermi gases the scanning gate could manipulate local modes like Andreev bound states \cite{Husmann:2015ab,Valtolina:2015ab}. It could also help to identify dynamical structures such as vortex patterns \cite{Beria:2013aa}.

\begin{acknowledgments}
We acknowledge discussions with A. Georges, T. Giamarchi, and thank B. Bräm, L. Corman, R. Desbuquois, R. Steinacher, P. Törmä, D. Weinmann and W. Zwerger for discussions and careful reading of the manuscript. We acknowledge financing from NCCR QSIT, the ERC project SQMS, the FP7 project SIQS, the Horizon2020 project QUIC, Swiss NSF under division II. SN acknowledges support from JSPS. JPB is supported by the Ambizione program of the Swiss NSF and by the Sandoz Family Foundation-Monique de Meuron program for Academic Promotion I.
\end{acknowledgments}

\clearpage

\makeatletter
\setcounter{section}{0}
\setcounter{subsection}{0}
\setcounter{figure}{0}
\setcounter{equation}{0}
\renewcommand{\theequation}{S\arabic{equation}}
\renewcommand{\thefigure}{S\arabic{figure}}
\renewcommand{\bibnumfmt}[1]{[S#1]}
\renewcommand{\citenumfont}[1]{S#1}

\section{Supplemental material}

\subsection{Experimental details}

\subsubsection{Experimental cycle}
To produce degenerate Fermi gases, we first create a mixture of the two lowest hyperfine states of $^6$Li and balance the populations using several incomplete Landau-Zener sweeps. Forced evaporation at a magnetic field of \SI{276}{G} cools the atoms to about the Fermi temperature. Subsequently, a complete Landau-Zener transition transfers the full population from the second to the third lowest hyperfine state, and the magnetic field is ramped to a Feshbach resonance at \SI{689}{G}. A magnetic field gradient induces the final evaporation and the magnetic field is ramped in \SI{200}{\milli\second} to \SI{949}{G}, setting the scattering length for transport to $-2653\,a_0$. The gas resides in a hybrid trap, where an optical dipole trap confines transversally ($x$, $z$), while the atoms are longitudinally ($y$) restricted by a magnetic field curvature. The trapping frequencies along $x$ and $z$ are \SI{224}{\Hz} and \SI{181}{\Hz} and in $y$ direction \SI{33.3}{\Hz}.

A chemical potential bias is induced by shifting the cloud in transport direction using a magnetic field gradient of about \SI{0.15}{G\per\centi\metre}. We then split the cloud asymmetrically with an elliptical repulsive beam into two reservoirs. By moving the magnetic trap back to its symmetric position, the atom number difference translates into a chemical potential bias. In the presence of the QPC the transport process is started by removing the repulsive beam and terminated by switching it back on again after a transport time of \SI{4}{\second}. Finally, we infer the density distribution using absorption imaging along the $x$ direction.

\subsubsection{Calibration of the quantum point contact}
The QPC provides harmonic confinement along the transverse directions ($x$, $z$) that we calibrate as follows. We measure the transverse trapping frequency $\omega_x$ by parametrically modulating the intensity of the laser beam restricting the atoms in $x$ direction. Close to the resonance the atoms are heated and escape from a dipole trap created by a laser beam that propagates along the $z$ axis and tightly confines the atoms to the center.

Then, we calibrate the frequency $\omega_z$ by measuring transport in the quantized regime, where the conductance plateaus energetically shift with the transverse confinement. Hence, a change in confinement along the $z$ axis can be compensated with the one along $x$, linking the unknown frequency to the calibrated one. Practically, we relate the two frequencies indirectly via calibrating each of them against a red-detuned beam that controls the chemical potential in the vicinity of the QPC. As a byproduct this beam is calibrated, which is relevant to characterize the scanning gate.

\subsubsection{Calibration of the scanning gate}
The strength of the scanning gate is extracted using transport. To this end, we center the gate on the channel and balance the added repulsion with a red-detuned beam (see red beam in figure 1c), previously calibrated. As the gate is tightly focused it easily blocks transport even at the highest achievable local chemical potentials. Hence, we enlarge the gate to waists of $w_x = \SI{1.3}{\micro\metre}$ and $w_y = \SI{2.0}{\micro\metre}$ to distribute the power and reduce the peak intensity.

Simultaneously, we image the scanning gate with a high-resolution microscope on a CCD sensor. Thanks to the measured local chemical potential shift the registered number of counts per exposure time is converted to the potential created on the atoms. With this the potential strength of the narrow scanning gate can be read from an image.

\subsubsection{Holographic beam shaping}
We operate a digital mirror device (DMD DLP5500 .55'' XGA from Texas Instruments) to create and move the scanning gate beam, while compensating for optical aberrations. It consists of a grid of micrometer sized square mirrors, which can be individually flipped about their diagonal axis to two stable positions (ON and OFF) at an angle of \SI{\pm12}{\degree}.

When all mirrors have the same orientation coherent light is diffracted into several orders. We send a collimated beam at \SI{532}{\nano\metre} onto the DMD incident at an angle close to \SI{-12}{\degree}, where the Littrow and the blazing condition are both fulfilled for the $-6^\text{th}$ order. In this configuration the diffraction and incident angles are identical and almost aligned with the specular reflection. This leads to a compact optical setup and a near-optimal diffraction efficiency of \SI{30}{\percent}. The pattern imprinted on the DMD is later imaged on the back focal plane of a microscope objective, which effectively projects its Fourier transform onto the atomic plane.

To correct aberrations and shape the beam, we control its phase locally by an amplitude hologram. We display a grating where groups of several consecutive mirrors in ON respectively OFF state are alternating. By shifting the grating locally we modify the phase of the light \cite{Zupancic2016}. The displayed grating has a larger spacing and hence additional diffraction orders appear. We use one directly neighboring the main $-6^\text{th}$ order, whose orientation relative to the hologram plane is controlled by the spacing and direction of the pixel grating. The orientation of this order then directly translates to its position in the atomic plane.

An identical microscope objective is placed symmetrically after the atomic plane to image the light potential. Cropping small apertures onto the DMD and fitting the position of their images through the optical system allows to retrieve the local tilt of the wavefront, similar to a Hartmann-Shack analysis. The tilt information is integrated to obtain a transverse spatial map of the beam phase, which is then compensated by distorting the lines of the initial grating to eliminate beam aberrations. We estimate a residual wavefront distortion of around $\lambda/10$.

\subsection{Data analysis}

\subsubsection{Temperature extraction}
For the temperature measurement, we ramp the magnetic field to a Feshbach resonance at \SI{689}{G} and image the cloud. With the virial theorem valid at unitarity \cite{Thomas2005} we determine the energy per particle from the second moment of the density distribution and translate it to entropy per particle by means of the known equation of state \cite{Ku2012, Guajardo2013}. To trace back the temperature to the BCS regime, we assume that the magnetic field is swept adiabatically \cite{Krinner2016}. Hence, the entropy at unitarity equals the one in the BCS regime, which reads for a weakly interacting, degenerate Fermi gas \cite{Carr2004, Su2003} 
\begin{equation}
	\frac{S}{N} = k_\text{B} \pi^2 \frac{T}{T_\text{F}} \left( 1 + \frac{64}{35 \pi^2} k_\text{F} a \right), 
	\label{eq:entropy}
\end{equation}
with the Fermi temperature $T_\text{F} = \hbar \bar{\omega} (6 N)^{1/3} / k_\text{B}$, the mean trapping frequency $\bar{\omega} = (\omega_x \omega_y \omega_z)^{1/3}$ and the corresponding wavevector $k_\text{F}$. With an interaction strength of $1/k_\text{F}a = -2.1$ the lowest order correction yields a value $T/T_\text{F}$ that is by \SI{9}{\percent} larger than the non-interacting case, supporting the expansion in $k_\text{F} a$. We obtain typical temperatures around \SI{60}{\nano\kelvin}. 

At these temperatures we expect our gas to be in the normal phase, as shown in the following. The critical temperature for superfluidity is locally largest at the center of the trap, where additionally an attractive beam increases the density. As a result the local Fermi temperature $\tilde{T}_\text{F}$ is about \SI{1.2}{\micro\kelvin} at our parameters, giving a local interaction strength of $1/\tilde{k}_\text{F} a = -1.3$. Thus, BCS theory including Gorkov and Melik-Barkhudarov corrections \cite{Pethick2002}, $T_c / \tilde{T}_\text{F} = 0.28 \exp [\pi / (2 \tilde{k}_\text{F} a)]$ predicts a critical temperature $T_c$ of \SI{42}{\nano\kelvin}. This is an upper bound as we neglect the presence of the repulsive beams forming the wire.

\subsubsection{Chemical potential bias extraction}
We infer the chemical potential bias initially prepared between the two reservoirs from the trapping geometry, the number of atoms, the temperature and the interaction. In each reservoir the potential is harmonic along the $x$ and $z$ direction and half-harmonic in the $y$ direction. Assuming a weakly interacting and degenerate Fermi gas in the left reservoir, the chemical potential reads \cite{Su2003}
\begin{equation}
	\mu_\text{L} = k_\text{B} T_\text{F,L} \left( 1 - \frac{\pi^2}{3} \left( \frac{T}{T_\text{F,L}} \right)^2 + \frac{512}{315 \pi^2} k_\text{F,L} a \right), 
	\label{eq:chemical_potential}
\end{equation}
with the Fermi temperature $T_\text{F,L} = \hbar \bar{\omega} (2 \cdot 6 N_\text{L})^{1/3} / k_\text{B}$, where the factor of two considers the half-harmonic confinement, and the corresponding wavevector $k_\text{F,L}$. Analogously, the formula holds for the right reservoir and the bias is given by $\Delta \mu = \mu_\text{L} - \mu_\text{R}$ and is typically $\SI{100}{\nano\kelvin} \cdot k_\text{B}$. The correction terms for finite temperature and interaction reduce the chemical potential each by about \SI{8}{\percent} compared to the non-interacting case.

\subsubsection{Conductance evaluation}
We infer the conductance from the exponentially decaying atom number difference $N_\text{L} - N_\text{R}$ between the left and right reservoir as described in \cite{Krinner2015}. Apart from the measured decay constant we need the compressibility $C$ of a single reservoir. It is calculated as $(\partial N_\text{L(R)} / \partial \mu_\text{L(R)})_T$ based on the equation of state \eqref{eq:chemical_potential} for a weakly interacting, degenerate Fermi gas and is given by
\begin{equation}
	C = \frac{(6 N)^{2/3}}{4 \hbar \bar{\omega}} \left( 1 - \frac{\pi^2}{3} \left( \frac{T}{T_\text{F}} \right)^2 - \frac{256}{105 \pi^2} k_\text{F} a \right), 
\end{equation}
where the quantities are evaluated at the same number of atoms $N_\text{L} = N_\text{R} = N/2$. The Fermi temperature reads $T_\text{F} = \hbar \bar{\omega} (6 N)^{1/3} / k_\text{B}$ and the corresponding Fermi wavevector is denoted by $k_\text{F}$. The temperature correction term decreases the compressibility by \SI{6}{\percent}, while the term for interactions increases it by \SI{11}{\percent} compared to the non-interacting value.

\subsubsection{Full width at half maximum of the transverse scans}
From the transverse cuts in figure~3 we extract the full width at half maximum (FWHM) by fitting the profiles for each tip strength separately. As the profiles vary from bell-shaped to flat-top, we use a modified Gaussian function inspired by the eigenstates of a harmonic oscillator, that we expect to faithfully describe the profiles. It reads 
\begin{equation}
	G(x) = \mathrm{e}^{-\left( \frac{x - x_0}{w} \right)^2} \cdot \sum_{k=0}^n a_{2k} H_{2k} \left(\frac{x - x_0}{w} \right) + G_\infty, 
\end{equation}
with the transverse position $x$ of the tip and $x_0$ of the channel's center. The parameter $w$ measures the width of the Gaussian function that is modified by the Hermite polynomials $H_k$ and their coefficients $a_k$. As the profiles are symmetric about the channel, we only include the even polynomials up to order $2n$. The order is chosen as small as possible while still reproducing the curves well and ranges practically from zero to six. The offset $G_\infty$ indicates the conductance when the gate position tends to infinity.

We evaluate the FWHM and its uncertainty numerically from the fitted function. The FWHM is given by the width at half maximum of $(G_\infty + G(x_0))/2$, while the uncertainty is obtained by shifting the half maximum level by the averaged standard error over the profile.

\subsubsection{Numerical derivative of the transverse scans}
The spatial derivative shown in figure~4 is evaluated separately for each transverse profile (see figure~3) at fixed tip strength. We extract the local slope using a weighted linear fit over four consecutive points, while its abscissa is the mean abscissa over the window. The weight includes the averaged standard error over the profile and leads to the uncertainty in the slope. We then shift the window by one point and repeat the method.

\subsubsection{Conductance noise estimation}
Generally, the measured conductances contain noise of different types and timescales as well as drifts that we need to separate. For example, white noise may be reduced upon averaging, while drifts worsen the result the more we average. To determine the minimal uncertainty, we perform an Allan analysis. This technique was initially developed to assess the stability of atomic clocks \cite{Allan1966} and is more widely used nowadays, for example to characterize optical tweezers \cite{Czerwinski2009}.

The basis of an Allan analysis is a time series of observations $y_0$, $y_1$, \dots, $y_{M-1}$ equally spaced by the measurement interval $\tau_0$. The Allan variance is calculated as \cite{Howe1981, Riley2008}
\begin{equation}
	s_m^2 = \frac{1}{2 (M - 2m + 1)}
	\sum_{k=0}^{M-2m} (\bar{y}_{k+m} - \bar{y}_k)^2, 
\end{equation}
where $\bar{y}_k = \frac{1}{m} \sum_{i=k}^{k+m-1}{y_i}$ are the observations averaged over $m$ consecutive points \footnote{Formula (10) in \cite{Riley2008} correctly gives the overlapping Allan variance, while formula (9) is wrong. The full inner sum should be squared and not just its summands.}. The bins are overlapping to obtain the best estimate of the Allan variance and it is equal to the classical version if the data are random and uncorrelated. Similarly, the Allan variance follows a chi-squared distribution that allows to estimate the confidence in the result. The degrees of freedom of the distribution is reduced due to the correlations among the bins and empirically given by \cite[equation (6.6)]{Howe1981}.

\begin{figure}
    \includegraphics{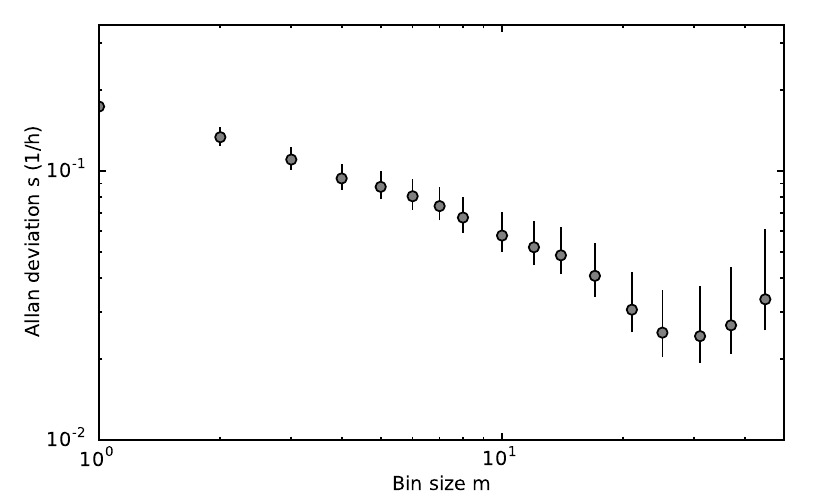}
    \caption{Allan analysis of conductances in the single mode regime. The analysis is based on 128 repeated measurements separated by a sampling time $\tau_0$ of \SI{24.3}{\second}. Two consecutive measurements that do not contain any atomic signal are filled with the last known conductance value. The error bars indicate the confidence intervals at a level of \SI{68.3}{\percent}.}
    \label{fig:allan_analysis}
\end{figure}

To assess the noise present in the conductance, we performed 128 repeated measurements in the single mode regime in the absence of the scanning gate. Figure~\ref{fig:allan_analysis} presents the Allan deviation of the conductance as a function of the bin size~$m$. With increasing bin size the deviation first decreases down to $0.024/h$ upon averaging 29 consecutive measurements, and then increases due to drifts in our apparatus. Although the minimal reachable noise of $0.024/h$ has asymmetric uncertainties, we conservatively assume the larger of the two limits, giving $0.024(12)/h$.

\subsection{Numerical simulations}
We use the Kwant package \cite{Groth2014} to compute the total transmission through the QPC in presence of a repulsive gate potential.
The quasi two-dimensional region separating the large atom reservoirs is discretized on a two-dimensional mesh covering an area $\SI{20}{\micro\metre} \times \SI{24}{\micro\metre}$ in the $x$-$y$ plane. The potential is simplified as the sum of the wire and scanning gate potential, since the attractive gate and optical dipole trap can be considered as uniform at the scale of the QPC.

The total transmission $\mathcal{T}(E)$ through the scattering region is then obtained by summing over all single-particle transport modes of the reservoirs along the transverse $x$-direction. We compute the conductance using a modified Landauer formula
\begin{equation}
	G = \frac{1}{\Delta\mu \cdot h} \sum_{n_z = 0}^\infty \int_{-\infty}^{+\infty} \mathcal{T}(E)  f(E + n_z \hbar \, \omega_z) \,\mathrm{d}E
	\label{eq:landauer_kwant}
\end{equation}
that now includes the contributions of every single mode $n_z$ in the transverse $z$-direction through an energy shift $n_z \hbar \, \omega_z$, assuming harmonic confinement along $z$. Each contribution to the total conductance is broadened by
\begin{equation}
	f(E) = \frac{1}{1 + \mathrm{e}^{(E - \mu_\text{L}) / k_\text{B} T}} - \frac{1}{1 + \mathrm{e}^{(E - \mu_\text{R}) / k_\text{B} T}}, 
\end{equation}
the difference between the Fermi-Dirac distributions at chemical potentials $\mu_\text{L} = \mu + \Delta\mu / 2$ and $\mu_\text{R} = \mu - \Delta\mu / 2$ and temperature $T$. 
Chemical potential is here defined as $\mu = E_F + V_G - \frac{1}{2} \hbar \omega_z$, the reservoir Fermi energy augmented by the attractive top-gate potential, and reduced by the zero-point energy along $z$.

In the end, we obtain scanning gate maps of the conductance by numerically performing the integral (\ref{eq:landauer_kwant}) for scattering potentials with different positions of the gate. The chemical potential imbalance $\Delta \mu$ and temperature $T$ used are extracted from time-of-flight pictures of the reservoirs, whereas the absolute chemical potential $\mu$ is obtained from a best fit to the experimental data.

\subsubsection{Fringe patterns in scanning gate maps}
In solid state samples scanning gate maps reveal fringe patterns, resulting from interferences between particles emitted by the point contact and reflected back by the scanning gate \cite{Topinka2001}. However, they are absent in our measured and simulated maps in figure~2. In this section we pinpoint the reason by studying the influence of averaging due to finite temperature and finite chemical potential bias.

\begin{figure}
    \includegraphics{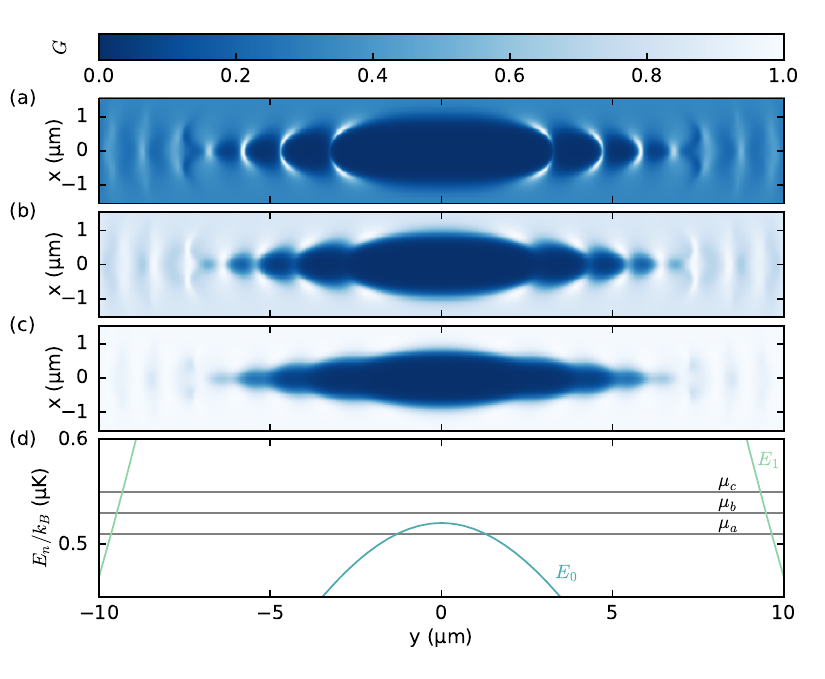}
    \caption{Simulated scanning gate maps at zero temperature and infinitesimal chemical potential bias. Panels (a), (b) and (c) show the conductance through the QPC as a function of the position of a local obstacle for three different chemical potentials: $\mu_\text{a} = \SI{510}{\nano\kelvin}$, $\mu_\text{b} = \SI{530}{\nano\kelvin}$ and $\mu_\text{c} = \SI{550}{\nano\kelvin}$. They are close to the zero-point energy of the QPC in transverse direction. The gate strength is common to all of them and is $\SI{620}{\nano\kelvin} \cdot k_\text{B}$, as in figure~2. (d) Effective potential along the transport axis, including the transverse mode energy of the beam defining the wire in $x$-direction. The colored curves correspond to the two lowest transverse mode energies labeled by $E_0$ and $E_1$. The horizontal lines (gray) indicate the chemical potentials used in the panels (a) to (c).}
    \label{fig:2dscan_noavg}
\end{figure}

As a reference, we study the case of zero temperature and infinitesimal bias, where the conductance reduces to $\mathcal{T}(\mu)/h$ for the first transverse mode in $z$-direction. Figure~\ref{fig:2dscan_noavg} presents scanning gate maps in this limit for three different chemical potentials, close to the transverse zero-point energy of the QPC. They are dressed by interferences, fading out as the gate is moving away. The fringes stem from multiple reflections between the gate and the QPC, reminescent of \cite{Topinka2001}. Some features of the pattern are understood with the aid of the effective potential in figure~\ref{fig:2dscan_noavg}d, which includes the transverse confinement in $x$-direction. As the particle moves outwards it gains kinetic energy and hence the phase of its wave function evolves faster, leading to a denser appearance of the resonances. Furthermore, they broaden as the transmission through the gate monotonically increases with higher kinetic energy.

\begin{figure}
    \includegraphics{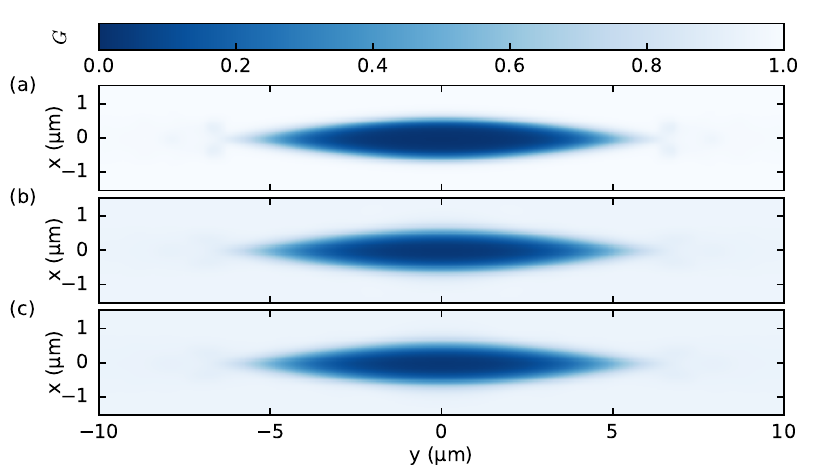}
    \caption{Simulated scanning gate maps including finite temperature and/or bias. The interferences are blurred in the case of finite bias only (a), finite temperature only (b) and the combined effect (c). A chemical potential of $\SI{688}{\nano\kelvin} \cdot k_\text{B}$, a bias of $\SI{99}{\nano\kelvin} \cdot k_\text{B}$ and a temperature of \SI{58}{\nano\kelvin} are used, as in figure~2b. The gate strength amounts to $\SI{620}{\nano\kelvin} \cdot k_\text{B}$.}
    \label{fig:2dscan_avg}
\end{figure}

The conductance at finite temperature, as well as at finite chemical potential bias is an average of the transmission for different particle energies. As the fringes move with the incident particle energy, they are blurred upon averaging. Taken separately, both finite bias and temperature are enough to wash out the interferences, shown in figure~\ref{fig:2dscan_avg}a and b. The combined effect can be seen in panel c.

\subsubsection{Transverse scans}
In figure~3 we study the transport of atoms through the channel as a function of the transverse position of the tip and its strength and interpret the experiment with an analytical model rooted in the Landauer formalism. Here, we investigate the numerical predictions based on the Kwant package and compare it to the results of the analytical model.

\begin{figure}
    \includegraphics{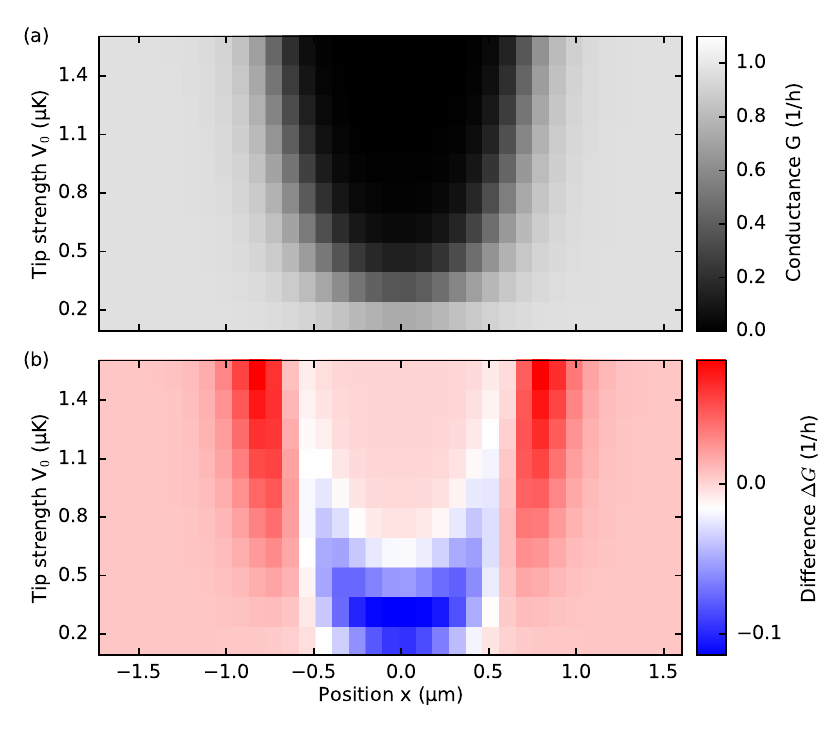}
    \caption{Transverse scan through the QPC center in the single mode regime, complementary to figure~3. (a) Numerically simulated conductance as a function of the transverse tip position~$x$ and strength~$V_0$. The simulation includes a single free parameter, the chemical potential in the reservoir, fitted to the measurement shown in figure~3b. The temperature and the chemical potential bias are \SI{58(5)}{\nano\kelvin} and $\SI{102(3)}{\nano\kelvin} \cdot k_\text{B}$ respectively, calibrated independently. (b) Difference of the conductance predicted by the analytical model, shown in figure~3c, and the numerical simulation.}
    \label{fig:transverse}
\end{figure}

Figure~\ref{fig:transverse} presents the predicted conductance as a function of the transverse tip position and its strength. Both models include the chemical potential as the only free parameter which is fitted against measured data, while the temperature and chemical potential bias are independently calibrated to \SI{58(5)}{\nano\kelvin} and $\SI{102(3)}{\nano\kelvin} \cdot k_\text{B}$. The optimized chemical potentials are $\SI{214}{\nano\kelvin} \cdot k_\text{B}$ for the analytical and $\SI{719}{\nano\kelvin} \cdot k_\text{B}$ for the numerical model. The reference of the chemical potential differs by the transverse mode energy $\hbar \omega_x/2$ of $\SI{521}{\nano\kelvin} \cdot k_\text{B}$ along the $x$ axis, in quantitative agreement with the fitted values. Additionally, the predicted conductances deviate at most by $0.1/h$ justifying the analytical model.

\subsection{Analytical model}
To analytically model the transverse scans presented in figure~3 we employ the Landauer formalism \cite{Ihn2010}. It describes the conductance through a one-dimensional channel connected to two reservoirs that are independently in thermal equilibrium, given for a single occupied transverse mode by
\begin{equation}
	G = \frac{1}{\Delta\mu \cdot h} \int_{-\infty}^{+\infty} \! \mathcal{T}(E) 
	\left[ f_\text{L}(E) - f_\text{R}(E) \right] \,\mathrm{d}E.
	\label{eq:landauer_model}
\end{equation}
Each reservoir is characterised by a Fermi-Dirac distribution~$f_\text{L}$ or $f_\text{R}$ respectively centred around the chemical potentials $\mu_\text{L}$ and $\mu_\text{R}$, biased by $\Delta\mu = \mu_\text{L} - \mu_\text{R}$ and with a temperature~$T$. The channel is incorporated in the energy-dependent transmission~$\mathcal{T}(E)$ that is dominated by the narrowest part of the constriction \cite{Glazman1988}.

At the narrowest place we neglect the extent of the transverse wave function compared to the potential created by the scanning gate tip. Hence the particles only explore a cut of the Gaussian tip potential along the transport direction ($y$)
\begin{equation}
	\mathcal{V}(x, y) = V_0 \mathrm{e}^{-2x^2/w_x^2} \cdot \mathrm{e}^{-2y^2/w_y^2}, 
\end{equation}
where $w_x$ and $w_y$ are the corresponding waists. To obtain an analytical expression for the transmission we approximate the Gaussian potential by a parabolic barrier
\begin{equation}
	\mathcal{V}(x, y) \approx V(x) - \frac{1}{2} m \Omega^2(x) y^2
\end{equation}
with a potential offset~$V(x)$ and anti-trapping frequency~$\Omega(x)$ depending on the transverse tip position~$x$.
\begin{align}
	V(x) = V_0 \mathrm{e}^{-2x^2/w_x^2} \\
	\Omega (x) = \frac{2}{w_y} \sqrt{\frac{V(x)}{m}} \label{eq:antitrap_freq}
\end{align}
The transmission through the parabolic barrier reads \cite{Kemble1935}
\begin{equation}
	\mathcal{T}(E) = \frac{1}{1+\mathrm{e}^{2 \pi \frac{E - V(x)}{\hbar \Omega (x)}}}, 
	\label{eq:parabolic_transmission}
\end{equation}
and we use it to approximate the transmission through the Gaussian barrier, valid for particles with energies above zero.

Finally, we calculate the conductance as a function of the transverse tip position~$x$ and strength~$V_0$ by numerically integrating the expression \eqref{eq:landauer_model} with the transmission through a parabolic barrier given in formula \eqref{eq:parabolic_transmission}. The chemical potential bias $\Delta\mu$ and the temperature $T$ are independently calibrated from absorption images, while the overall chemical potential~$\mu$ is a free parameter fitted to the experimental data in figure~3b.

\subsubsection{Validity of the parabolic approximation}
\begin{figure}
    \includegraphics{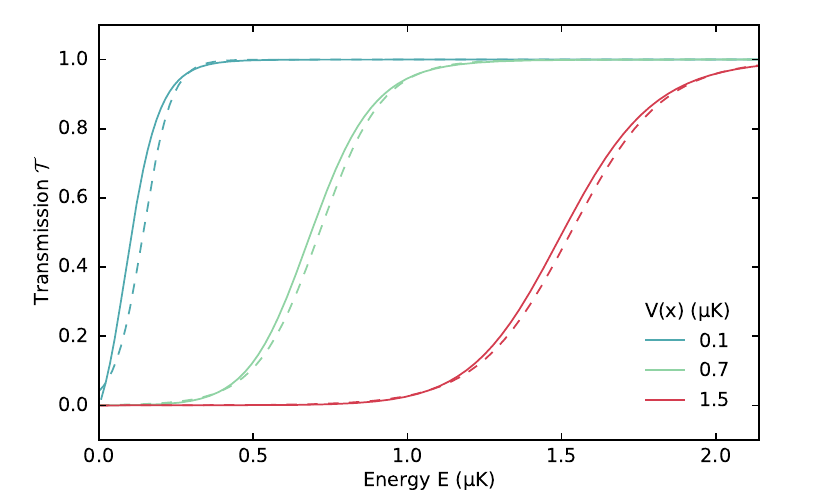}
    \caption{Energy-dependent transmission~$\mathcal{T}$ through a Gaussian barrier (solid line) and its parabolic approximation (dashed line) for different potential profiles, characterized by $V(x)$. In the parabolic case the transmission is calculated using equations~\eqref{eq:antitrap_freq} and \eqref{eq:parabolic_transmission}. For the Gaussian potential the one-dimensional, time-independent Schrödinger equation is numerically integrated using Numerov's method \cite{Hairer1993}. The parabolic approximation describes the transmission well within the experimental parameters.}
    \label{fig:transmissions}
\end{figure}

Figure~\ref{fig:transmissions} compares the transmission through a Gaussian and a parabolic barrier for various potential offsets~$V(x)$ and hence also anti-trapping frequencies. In the experiment the offset~$V(x)$ ranges from essentially zero, when the tip is weak and far apart from the QPC, up to \SI{1.5}{\micro\kelvin} for the strongest tip placed at the center. With increasing offset the transition from zero to unit transmission is shifted to higher particle energies. At the same time the anti-trapping frequency increases as the potential barrier gets sharper and the transition gets broader due to quantum tunnelling and reflections. The transmission through the parabolic barrier faithfully approximates the one through the Gaussian obstacle and only slightly underestimates it around the barrier top~$V(x)$. The deviations increase for lower barrier heights.

\end{document}